\documentclass[12pt]{iopart}
\usepackage{epsf}  

\begin{document}

\title{Perturbation theory for the effective diffusion constant in
a medium of random  scatterers}

\author{D.S. Dean\dag\ddag, I.T. Drummond\dag, R.R. Horgan\dag\   and 
A. Lef\`evre\S }

\address{\dag \ DAMTP, CMS, University of Cambridge, Cambridge, CB3
0WA, UK}

\address{\ddag\ IRSAMC, Laboratoire de Physique Th\'eorique,
Universit\'e Paul Sabatier, 118 route de Narbonne, 31062 Toulouse
Cedex 04, France}

\address{\S \ Rudolf Peierls Centre for Theoretical Physics, 
University of Oxford, 1 Keble Road, Oxford, OX1 3NP, UK}

\begin{abstract}
We develop perturbation theory and physically motivated
resummations of the perturbation theory for the
problem of a tracer particle diffusing in a random media. The
random media contains point scatterers of density $\rho$
uniformly distributed through out the material. The tracer
is a Langevin particle subjected to the quenched random force
generated by the scatterers. Via our perturbative analysis we
determine when the random potential can be approximated by a 
Gaussian random potential. 
We also develop a self-similar renormalisation
group approach based on thinning out the scatterers, this scheme is similar
to that used with success for diffusion in Gaussian random potentials and
agrees with known exact results.  To assess the accuracy of this 
approximation scheme its predictions  are confronted with 
results obtained by numerical simulation.

\end{abstract}



\section{Introduction}
The bulk transport properties of quenched random media play an
important role in engineering sciences such a hydrology \cite{hyd}.
Calculational techniques to evaluate bulk transport properties such as
effective diffusivities, mobilities, conductivities and permeabilities
thus have far reaching applications. In systems such as zeolites,
disorder modifies the diffusion of reacting chemical species and thus
plays a crucial role in diffusion limited chemical reactions
\cite{chem}.  The study of such systems is also relevant to the
evaluation of short-time transport properties in non-quenched random
media where the time scale on which a tracer diffuses over the typical
correlation scale of the advecting field is much shorter than the time
scale controlling the temporal evolution of the advecting
field. Systems with quenched disorder may also be used as paradigms
for systems with temporal disorder and may sometimes be used to obtain
bounds on the transport properties of non-quenched fields.

Recently there has been much interest in the nature of the glass
transition, in particular as to whether it is a purely dynamical or a
static transition. Consider a tracer particle diffusing and
interacting with other identical particles as could be the case in a
liquid or gas. If inertial effects are not important we can assume
that the motion of the particle is controlled by over-damped Langevin
dynamics. In the liquid, or ergodic, phase we expect the tracer
particle to have a non-zero self diffusion constant $\kappa$. Now
consider a situation in which the same particle diffuses in a quenched
background where all the other particles have been frozen in a
particular configuration, possibly chosen from Gibbs-Boltzmann
equilibrium ensemble. Let the diffusion constant in this quenched
system be $\kappa_{e}$. It has been shown \cite{maths} that
$\kappa_{e} < \kappa$; a result that seems physically plausible since,
if the background particles can move about, the cages which could trap
the tracer in the quenched background, will break up on some
time-scale and free it to disperse more quickly than in the quenched
case. This consideration recently lead Szamel \cite{szamel} to study
this problem as a test of the mode coupling theory for interacting
particle systems.

A Langevin particle diffusing in a random potential is often invoked
to describe, at a suitably coarse grained level, the transport
properties of certain random media. Often the first choice of
statistics for the potential is Gaussian.  If the potential  has a
finite correlation length, then we can subdivide the system into 
uniform blocks whose size is larger than
that correlation length. Consequently the diffusion processes within
each of these blocks will be independent of one another. The particle
will stay within each block for a random time $\tau$, that will be
distributed similarly but independently for each block. For Gaussian
statistics we expect the mean value of $\tau$ to be finite.   
Thus, viewed as coarse grained hopping process between blocks, 
the motion of the tracer will appear to be an uncorrelated random walk and thus
asymptotically exhibit normal diffusion.

The calculation of the
diffusion constant of a Langevin particle in a Gaussian potential,
specified therefore solely by its two point correlation function
$\Delta({\bf x})$, has been attempted via a wide variety of
approximation schemes \cite{rg}.  The most successful to date is the
self-similar renormalisation group (RG) scheme of \cite{rg}. Other
approximation schemes such as self-consistent perturbation theory,
and the Hartree-Fock approximation perform poorly when confronted
with known exact results and numerical simulations.
The modification of the diffusion constant in the presence of inertial
effects \cite{inertia} can also be well explained using a Gaussian
closure scheme, within the Martin-Siggia-Rose path integral
formalism \cite{msr}. This closure approximation  is rather 
reminiscent of certain aspects of mode-coupling theory \cite{mct} 
and is related to the dynamics of mean field spin glasses \cite{bckm}
(the approximation becomes exact in the limit of infinite
spatial dimension and when the field strength is suitably scaled). 
This Gaussian closure predicts a 
glass transition \cite{inertia} (signalled by a vanishing of the diffusion 
constant and ageing in correlation and response functions) for a Langevin 
process diffusing in a short range correlated, finite variance Gaussian random
potential, however from our previous discussion we do not expect 
such a transition in finite dimensions. 

In this paper we shall consider the diffusion constant of a Langevin
tracer particle diffusing in a system of quenched random scatterers
which are independently and uniformly dispersed through out the
volume.  This corresponds to the situation found in zeolites where the
scatterers are usually charged ionic sites.  We draw attention to the
fact that this model is distinct from that of the Lorentz gas in which
an undamped Newtonian particle is scattered by hard spheres in three
dimensions or hard discs in two dimensions (overlapping or
non-overlapping) \cite{lorentz}.

We develop a double expansion for the self diffusion constant in
interaction strength and the density of scatterers and determine the
range of validity of the Gaussian approximation for the random field 
statistics.  We then evaluate the
density expansion to first order and use renormalisation group ideas
to resum this expansion. This strategy leads to a result which agrees
with known exact results for diffusion the presence of scatterers in
one and two dimensions \cite{scat1}.  We identify the situations where
this approximation appears to be accurate by comparison with numerical
simulations of diffusion in the presence of hard and soft sphere
scatterers. A distinct trend appears from our study, a renormalization
group calculation based on the weak density expansion appears to 
work very well in problems where the scatterers attract the tracer particle,
but is considerably less performant where the scatterers are repulsive. 

A number of useful mathematical results for Langevin
particles diffusing in random potentials, which we shall make use of
in this paper, are given straightforward derivations in the appendix,
thus making the paper self contained.

\section{The model}
The position ${\bf X}_t$, of the Langevin tracer particle subject to a
white noise, $\eta(t)$ and a force generated by a potential $\phi({\bf
x})$, satisfies the stochastic differential equation
\begin{equation}
{d{\bf X}_t\over dt}=\sqrt{\kappa}{\bf \eta}(t) -\lambda
\nabla\phi({\bf X}_t)~~,
\label{eqlan}
\end{equation}
where
\begin{equation}
\langle \eta_i(t)\eta_j(t')\rangle=2 \delta_{ij}\delta(t-t')~~,
\end{equation}
and $\langle\cdots\rangle$ denotes an average over the white noise.
The Einstein relation implies that the local, or bare, diffusivity
$\kappa$ and the coupling to the potential gradient $\lambda$ are
related by the equation
\begin{equation}
\frac{\lambda}{\kappa}=\beta=\frac{1}{T}~~,
\end{equation}
where $T$ is the absolute temperature in appropriate units.  The
probability density, $p({\bf x},t)$, for the position of the particle
obeys
\begin{eqnarray}
{\partial p\over \partial t} &=& \kappa\nabla^2 p + \lambda\nabla
\cdot (p \nabla \phi) \\ &=& -H p~~,
\label{deg} 
\end{eqnarray}
where equation (\ref{deg}) defines the positive operator $H$.  The
effective diffusivity, $\kappa_e$, describing the dispersion of the
particle at large times and distances is defined by
\begin{equation}
\kappa_e=\frac{1}{2D}\lim_{t\rightarrow\infty}\frac{\langle {\bf
X}_t^2\rangle}{t}~~,
\end{equation}
where $D$ is the dimension of space. The mean squared displacement of
the particle is calculated via
\begin{equation}
\langle {\bf X}_t^2\rangle=\int d{\bf x\ x}^2 p({\bf x},t)~~.
\end{equation}

In this paper we will consider potentials of the type
\begin{equation}
\phi({\bf x})=\sum_{n=1}^{N} V({\bf x}-{\bf x}_n)~~,
\end{equation}
generated by $N$ scatterers frozen in the volume ${\cal V}$.  Here ${\bf
x}_n$ is the position of scatterer $n$ and in what follows all the
${\bf x}_n$ will be taken to be independently and uniformly
distributed in the volume ${\cal V}$. This situation where the
scatterers are uniformly and independently distributed could arise in
circumstances where the medium is formed by the rapid quench of a very
high temperature liquid or gas to a solid phase, the scatterers then
becoming frozen in the high temperature configuration.  The density of
scatterers is $\rho = N/{\cal V}$~. The thermodynamic limit is
${\cal V}\rightarrow\infty$ at fixed $\rho$~.

The potential $V({\bf x})$ for a scattering particle at the origin
will be taken to be rotationally invariant that is $V({\bf x}) =
V(|{\bf x}|)$.  The quenched disorder in this problem is therefore
completely specified by the positions, ${\bf x}_n$, of the scatterers
and the disorder average is given by
\begin{equation}
\langle {\cal O} \rangle_d =\prod_{n} \frac{1}{\cal V}\int d{\bf
x}_n~~{\cal O}~~. \label{uni1}
\end{equation}

The Green's function for the problem is $G'({\bf x},{\bf x'}) =
H^{-1}({\bf x},{\bf x}')$.  It satisfies the equation
\begin{equation}
\kappa\nabla^2 G' + \lambda\nabla \cdot (G' \nabla \phi) =
-\delta({\bf x-x}')~~.
\label{eqG}
\end{equation}
The Fourier transform ${\tilde G}'({\bf k}$, may be defined to be
\begin{equation}
{\tilde G'}({\bf k}) = \int d{\bf y}\ \exp(-i{\bf k}\cdot {\bf y})
G'({\bf y+x}',{\bf x}')~~.
\end{equation}
Because the system is statistically homogeneous in space we may for
most purposes set ${\bf x}'=0$ in the above without loss of
generality.  From equation (\ref{eqG}) we have in Fourier space
\begin{equation}
{\tilde G'}({\bf k}) = {1\over \kappa {\bf k}^2} - {\lambda\over
\kappa {\bf k}^2}\int {d{\bf q}\over (2\pi)^D} {\bf k}\cdot {\bf q}\
{\tilde \phi}({\bf q}) {\tilde G'}({\bf k}-{\bf q})~~.
\label{eqg1}
\end{equation}
The effective diffusion constant $\kappa_e$ can now be obtained from
the long distance behaviour of the Green's function \cite{rg}~.  In
Fourier space we have
\begin{equation}
{\tilde G'}({\bf k}) \sim {1\over \kappa_e {\bf k}^2}\ \ \rm{as}\
|{\bf k}|\to 0~~.
\end{equation}

One can calculate $\kappa_e$ exactly for this problem in one dimension
\cite{scat1} and one finds
\begin{equation}
\kappa_e = \kappa\exp\left(-2\rho \int dx\ \left( \cosh(\beta V(x)) -1
\right)\right)~~.\label{oned}
\end{equation}
Remarkably the expression (\ref{oned}) is independent of the sign of
$\beta$. For positive $\beta$ the particle is attracted to the minima
of the potential $\phi$ and for negative $\beta$ it is attracted to
the maxima. To disperse however, the particle must visit both minima
and maxima, due to the absence of saddles in one dimension, and so one
could argue that, in some sense, both minima and maxima were equally
important for determining the dispersion. The result is still however
rather intriguing from a physical point of view.  Another interesting
property of the one dimensional result is that if one considers a
system with two types of scatters, types $a$ and $b$, one type of
density $\rho_a$ with potential $V_a$ and the other of density
$\rho_b$ with potential $V_b$ one finds that
\begin{equation}
\kappa_{e}(a+b )= {\kappa_e(a)\kappa_e(b)\over \kappa}~~, \label{mult}
\end{equation}
where $\kappa_{e}(a+b)$ is the effective diffusivity for system with
both scatterer types and $\kappa_e(a)$ and $\kappa_e(b)$ are the
effective diffusivities for the system with only particles of type $a$
and $b$ respectively.

An exact result was also recently found in two-dimensions. If one
considers a system of charged scatterers where
\begin{equation}
\phi(x)=\sum_{n=1}^{N}q_n V({\bf x}-{\bf x}_n)~~,
\end{equation}
and where $q_n$ is the charge of the scatterer at $x_n$ which takes
the (quenched) value $q_n = \pm 1$ with probability $1/2$ so that the
system is statistically electro-neutral, then it was shown that
\begin{equation}
\kappa_e = \kappa \exp\left(-\rho\int d{\bf x} \left[\cosh\left(\beta
V({\bf x})\right) -1\right]\right)~~.\label{twod}
\end{equation}
The interesting feature in this case is again that the multiplicative
structure seen in equation (\ref{mult}) in the case of two families of
independent charged scatterers still holds.

\section{Weak disorder expansion}
\begin{figure}
\begin{center}
\epsfxsize=0.7\hsize\epsfbox{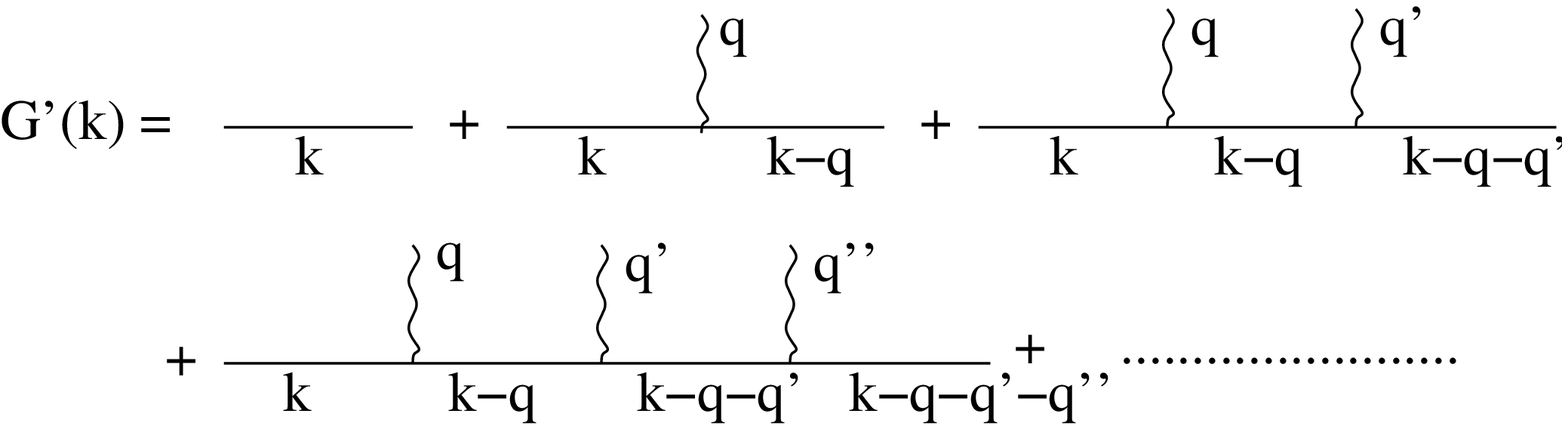}
\end{center}
\caption{\label{g1}Diagrammatic expansion for ${\tilde G}'({\bf k})$}
\end{figure}

In the non-neutral case in two and higher dimensions one must revert
to approximate methods to calculate $\kappa_e$.  We consider the
perturbation expansion generated by iterating equation (\ref{eqg1})
and which is shown diagrammatically in figure (\ref{g1}).

\begin{figure}
\begin{center}
\epsfxsize=0.7\hsize\epsfbox{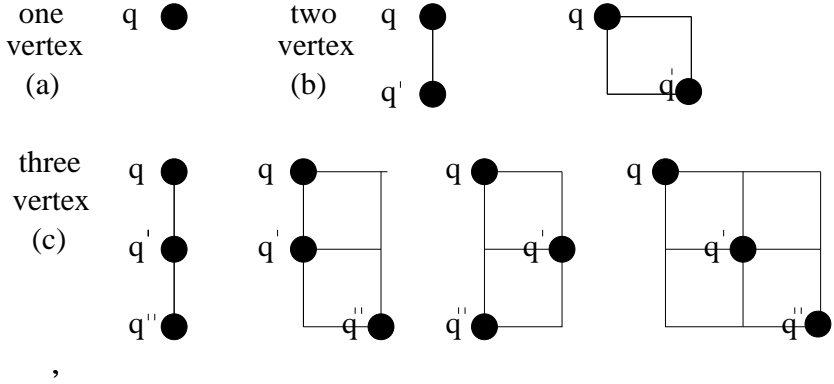}
\end{center}
\caption{\label{b1}Blob diagrams to order $\lambda^3$}
\end{figure}

The Feynman rules for this perturbation theory (before averaging over
the disorder) are:
\begin{itemize}
\item The total momentum at each vertex is conserved.
\item Each solid horizontal line with momentum $\bf k$ flowing through
it is the bare propagator ${1/\kappa {\bf k^2}}$.
\item Each wavy line with momentum ${\bf q}$ flowing into it carries a
factor $-\lambda {\bf k}\cdot {\bf q}{\tilde \phi}({\bf q}) $ where
$\bf k$ is the incoming (from the horizontal left) momentum.
\item Each wavy line momentum ${\bf q}$ in the diagram is integrated
with the measure $d{\bf q}/(2\pi)^D$.
\end{itemize}
One now has to average over the disorder to obtain
\begin{equation}
G(k) = \langle{\tilde G'}({\bf k})\rangle_d~~,
\end{equation}
where $k =|{\bf k}|$, as the averaged Green's function is clearly
isotropic.  We now note that the Fourier transform of $\phi$ is given
by
\begin{equation}
{\tilde \phi}({\bf q}) = \sum_{n}{\tilde V}({\bf q})\exp(-i{\bf
q}\cdot {\bf x}_n)~~.
\end{equation}
Therefore each vertex contains $n$ terms of the form ${\tilde V}({\bf
q})\exp(-i{\bf q}\cdot {\bf x}_n)$. If one considers a diagram coming
from the expansion of all the sums at each vertex containing the
particle index $i$, $k$ times at the vertices $\nu_1 \cdots \nu_k$,
then the contribution depending on the position ${\bf x}_i$ of
scatterer $i$ is given by the prefactor
\begin{equation}
P = \exp(-i\sum_{m=1}^k {\bf x}_i\cdot{\bf q}_{\nu_m})~~.
\end{equation}
Now, the disorder average with respect to ${\bf x}_i$ yields
\begin{equation} 
\langle P \rangle_d = {(2\pi)^D\over\cal V}\delta(\sum_{m=1}^k {\bf
q}_{\nu_m})~~.
\label{cons}
\end{equation}
Hence the total momentum flowing up into vertices with the same
scatterer index is zero. We also note that in the unaveraged Feynman
rules any vertex with upward flowing momentum ${\bf q}_\nu =0$ causes
the diagram containing it to be zero. In each diagram one has $N$
choices for the scatterer index at the first vertex. Every distinct
scatterer in each diagram carries a factor of $1/{\cal V}$ from
averaging over its position.  A diagram with $r$ distinct scatterer
indices at its vertices (but with the positions of each index fixed)
therefore carries a factor $\rho^r$ because there is a factor $N^r$
for the number of ways to choose the scatterers plus a factor of
$1/{\cal V}^r$ from the averaging over the $r$ distinct scatterer
positions.  We can represent this ordering graphically as is shown in
figure (\ref{b1}) where the one, two and three vertex contributions
are shown, the first column represents the first scatterer chosen in
the first vertex, the momentum at each blob is the momentum flowing
into that vertex. We continue this development where the $\nu^{th}$
row corresponds to the $\nu^{th}$ vertex.  In each of these diagrams
the sum of the momentum down each column is zero due to equation
(\ref{cons}).  Also, from the Feynman rules, any diagram with a blob
containing zero momentum must be zero; along with momentum
conservation this means that any non-zero diagram must contain each
scatterer zero times or at least twice.  The diagram figure
(\ref{b1}a) is thus zero as are all but the first of the diagrams in
figures (\ref{b1}b) and (\ref{b1}c).
The order of each diagram written this way is $\rho^c\times
\lambda^r$, where $c$ is the number of columns,corresponding  to
distinct particles, and $r$ is the number of rows, corresponding to the
number of vertices.

\begin{figure}
\begin{center}
\epsfxsize=0.7\hsize\epsfbox{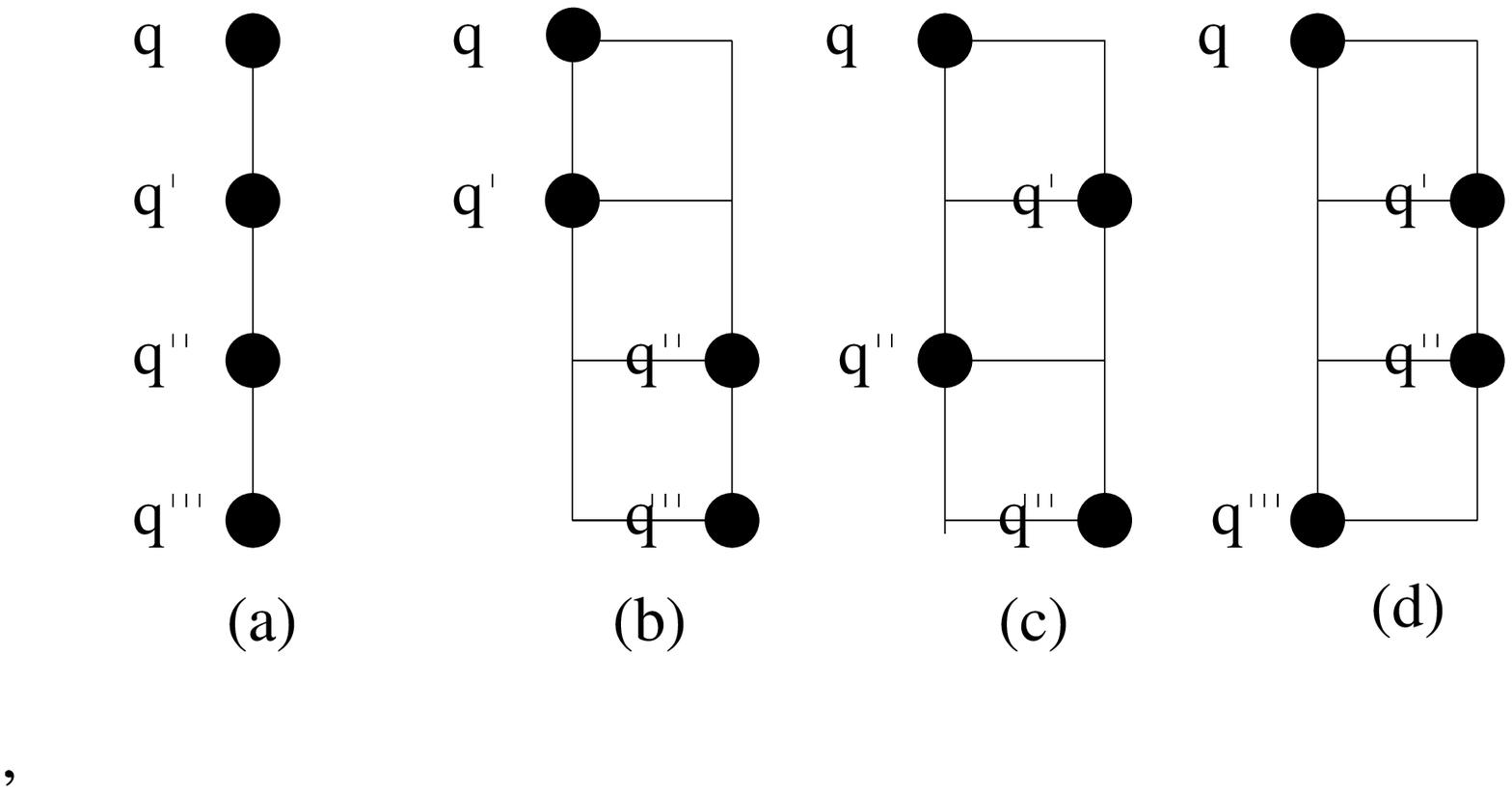}
\end{center}
\caption{\label{b2}Non-zero blob diagrams of order $\lambda^4$}
\end{figure}

In figure (\ref{b2}) we show the non-zero four-vertex contributions:
figure(\ref{b2}a) has only one column (scatterer) and figures
(\ref{b2}b), (\ref{b2}c) and (\ref{b2}d) represent the two scatterer
diagrams.

We note that for large $\rho$ the dominant diagrams with an even
number $2n$ of rows are those where each column has only two blobs and
that this diagram, in units where $\kappa =1$, is of order
$(\rho\lambda^2)^n$. Note that these diagrams are formed by pairing
the vertices and reproduce the diagrams for a particle diffusing in a
Gaussian random field \cite{rg}, where the field $\psi$ has the
correlation function given by
 
\begin{equation}
\langle \psi({\bf q})\psi({\bf q'})\rangle_d = (2\pi)^{D}\rho\lambda^2
\delta({\bf q}+ {\bf q'}){\tilde V}({\bf q}) {\tilde V}(-{\bf q})~~.
\end{equation}

For diagrams with an odd number of vertices $2n+1$ the dominant
diagrams in $\rho$ are those with $n-1$ pairs of blobs with a single
column containing three blobs and the order of these diagrams is, in
units where $\kappa =1$, $\rho^{n-1}\lambda^{2n+1} = (\rho\lambda^2)^n
\lambda/\rho$. Thus in the limit $\lambda \to 0$ whilst keeping
$\rho\lambda^2 =$ constant, one recovers the limit where a system of
scatterers generates a potential which can be treated as Gaussian for
the purposes of calculating $\kappa_e$.  This limit can also be seen
in exactly soluble one dimensional and two dimensional models
\cite{scat1}.

The modified Feynman rules leading to systematic expansion of $G({\bf
k})$ in terms of $\lambda$ are thus:
 
\begin{itemize}
\item At order $\lambda^r$ draw all distinct blob diagrams with $r$
rows.
\item Each blob diagram carries a factor of $\rho^c$ where c is the
number of columns.
\item Each blob diagram has a corresponding Feynman diagram with the
same rules as before except it is ${\tilde V}({\bf q})$ at each vertex
instead of ${\tilde\phi}({\bf q})$.
\item The total momentum down each line is zero and each independent
remaining momentum is integrated with the measure $d{\bf q}/(2\pi)^D$.
\end{itemize}

Finally the series can be Dyson resummed in terms of one-particle
irreducible diagrams by writing
\begin{equation}
G(k) = {1\over \kappa k^2 - \Sigma( k)}~~,
\end{equation}
where $\Sigma$ is given by the summed one particle irreducible blob
diagrams multiplied by $\kappa^2 k^4$, that is to say the
corresponding diagram with the first in going and the last out going
bare propagators amputated. As usual a one-particle irreducible
diagram is one that does not factorise. All the diagrams of order
$\rho$ are clearly one-particle irreducible. In figure (\ref{b2}),
figure (\ref{b2}b) is clearly reducible and the others are
irreducible.

In terms of the one particle irreducible diagrams
\begin{equation}
\kappa_e = \kappa -\lim_{k \to 0}{ \Sigma(k)\over k^2}~~.
\end{equation}

In what follows we shall use the fact that ${\tilde V}({\bf q}) =
{\tilde V}(q)$ where $q = |{\bf q}|$, due to the radial symmetry the
the potential $V$.  Using the perturbation theory rules developed here
we find that to order $\beta^3$
\begin{eqnarray}
\fl {\kappa_e\over \kappa} &=& 1 - {\rho\beta^2\over D}\int {d{\bf
q}\over (2\pi)^D} \ {\tilde V}^2(q) + {\rho\beta^3\over D} \int {d{\bf
q}\over (2\pi)^D} {d{\bf p}\over (2\pi)^D} \left(1 - {({\bf
p}\cdot{\bf q})^2\over p^2 q^2}\right){\tilde V}(q) {\tilde
V}(q){\tilde V}(|{\bf p} + {\bf q}|)\nonumber \\ \fl &+&
{\rho^2\beta^4\over 2 D^2} \left[\int {d{\bf q}\over (2\pi)^D} \
{\tilde V}^2(q)\right]^2 + \beta^4\rho R~~.
\label{eqb4}
\end{eqnarray}
The principal difficulty with the above perturbation theory is the
identification of the $O(k^2)$ term in each one-particle irreducible
diagram. The term $R$ not written explicitly in the above suffers from
this problem and it is not clear how the resulting integral should be
expressed for efficient evaluation.  Also expansion as a power series
in $\beta$ is presumably not very accurate where $\kappa_e$ deviates
significantly from its bare value, as the first term is $O(\beta^2)$.
\section{Weak density expansion}
One of the principal technical difficulties with the weak disorder
expansion as presented in the previous section is that its convergence
can be expected to be poor, and it may possess divergences for
singular potentials. In addition, as the interested reader may verify,
the identification of $\kappa_e$ by extracting the terms $O(k^2)$ in
the expansion of $\Sigma(k)$ is algebraically rather difficult.  In
the case of scatterers it is natural to examine the expansion of
$\kappa_e$ in terms of the density $\rho$ of the scatterers.  The
small density expansion was pioneered by Maxwell \cite{max} in the
context of effective dielectric constants. We shall develop this
analysis for the problem of the scatterers in a compact and direct
manner. We start with the formal expression for the diffusion constant
which is given in terms of the Green's function by
\begin{equation}
{\kappa_e\over \kappa} = 1 - {\kappa\beta^2\over D Z}\int d{\bf x}
d{\bf x}' \ G'({\bf x},{\bf x}') \nabla \phi({\bf x}) \cdot
\nabla\phi({\bf x}') \exp\left(-\beta \phi({\bf x}')\right)~~,
\label{krep}
\end{equation}
where $Z$ is the one particle partition function for a particle in the
potential $\phi$ given by
\begin{equation}
Z = \int_{\cal V} d{\bf x} \exp\left(-\beta \phi(\bf x)\right)~~.
\end{equation}

This result does not appear to be widely known (or at least used) in
the physics literature. It can be derived from static considerations
\cite{der} or via the stochastic differential equation for the
diffusion process \cite{spohn}. For a proof based on the stochastic
approach we refer the reader to the appendix of this paper.

The expression equation (\ref{krep}) may be used to reformulate the
weak disorder expansion by exploiting the perturbation expansion for
$G'$. We see that to order $\rho$ the effective diffusion constant is
the result of independent contributions from $N$ scatterers, that is
\begin{equation}
\fl {\kappa_e^{(1)}\over \kappa} = 1 - \sum_n {\kappa\beta^2\over D
Z_{1,n}}\int d{\bf x} d{\bf x}' \ G_1({\bf x},{\bf x}', {\bf x}_n)
\nabla \phi_{1,n}({\bf x}) \nabla\phi_{1,n}({\bf x}') \exp\left(-\beta
\phi_{1,n}({\bf x}')\right)~~, \label{ke1a}
\end{equation}
where $\kappa_e^{(1)}$ denotes $\kappa_e$ calculated to first order in
$\rho$.  In the above $\phi_{1,n}({\bf x}) = V({\bf x} -{\bf x}_n)$ is
the potential due to the scatterer at ${\bf x}_n$ and $Z_{1,n}$ the
corresponding one particle partition function. We have denoted by
$G_1({\bf x},{\bf x}', {\bf x}_n) $ the diffusion Green's function for
a tracer particle in the presence of only one scatterer at ${\bf
x}_n$. Note that without loss of generality we may take all the ${\bf
x}_n = 0$ and in the limit of large ${\cal V}$ and $N$, when $V({\bf x})$
goes to zero sufficiently quickly as $|{\bf x}| \to \infty$ so that
the integral
\begin{equation}
F = \int d{\bf x}' \left[1- \exp\left(-\beta V({\bf x}')\right)\right]
\label{intf}
\end{equation}
is finite, then
\begin{equation}
{\kappa_e^{(1)}\over \kappa} = 1 - \rho {\kappa\beta^2\over D}\int
d{\bf x} d{\bf x}' \ G_1({\bf x},{\bf x}', 0) \nabla V({\bf x})
\cdot\nabla V({\bf x}') \exp\left(-\beta V({\bf
x}')\right)~~. \label{ke1b}
\end{equation}
We note here that to order $\rho$ the value of $\kappa_e$ is
independent of the nature of the correlations between the ${\bf x}_n$
and the formula equation (\ref{ke1b}) relies only on the fact that the
distribution of scatterers is homogeneous and isotropic.  We now
define
\begin{equation}
{\bf w}({\bf x}) = \beta \kappa \int d{\bf x}' G_1({\bf x},{\bf x}',
0) \nabla V({\bf x}') \exp\left(-\beta V({\bf x}')\right)~~,
\end{equation}
and by the definition of $G_1$, the Green's function for diffusion in
the presence of a single scatterer taken to be at the origin, we have
that
\begin{equation}
\nabla^2 {\bf w}_i({\bf x}) + \beta \nabla \cdot {\bf w}_i({\bf x})
\nabla V({\bf x}) = -\beta\nabla_i V({\bf x}) \exp\left(-\beta V({\bf
x} )\right)~~.
\label{eqw}
\end{equation}
Using the spherical symmetry of the problem we write $ {\bf w} =
\exp(-\beta V(r)) g(r) {\bf x}/r$ where $r= |{\bf x}|$ and $g$ obeys
\begin{equation}
 {d^2\over dr^2} g + {D-1\over r}{d\over dr}g - {D-1\over r^2}g -
\beta{d\over dr}g \ {d\over dr}V + \beta{d\over dr}V=0 ~~.\label{eqg}
\end{equation}
In terms of $g$ we thus have
\begin{equation}
{\kappa_e^{(1)}\over \kappa} = 1 + {\rho S_D\over D} \int_0^\infty
r^{D-1} dr\ \left[{\partial\over \partial r} \exp\left(-\beta V(r)
\right)\right] g(r) \label{smallrho}~~,
\end{equation}
where $S_D$ is the area of a unit sphere in $D$ dimensions. The
behaviour of $g$ at large $r$ needs to be determined. Note that if $V$
decays sufficiently quickly at large $r$ then we can write equation
(\ref{eqw}) as
 \begin{equation}
\nabla^2 {\bf w}_i({\bf x}) \approx -\beta\nabla_i V({\bf x})
\exp\left(-\beta V({\bf x} )\right)~~,
\label{eqw1}
\end{equation}
for large $r$. The effect of the source term must die away as $r \to
\infty$ for rapidly decaying potentials and so at large $r$ we may
write ${\bf w} = \nabla \psi$ where $\psi$ obeys
\begin{equation}
\nabla^2 \psi = \exp\left(-\beta V\right) - 1~~,
\end{equation}
which has the solution
\begin{equation}
\psi({\bf x}) = \int d{\bf x}' \Delta^{-1}_D({\bf x},{\bf x'})
\left[1- \exp\left(-\beta V({\bf x}')\right)\right]~~,
\end{equation}
where
\begin{equation}
\Delta^{-1} _D= {C_D\over |{\bf x}-{\bf x}'|^{D-2}}~~.
\end{equation}
is the inverse of the Laplacian in $D$ dimensions (the case $D=2$ is
to be interpreted as a logarithm).  Now at large $|{\bf x}|$
\begin{equation}
{\bf w}({\bf x})\approx \nabla\psi({\bf x}) \approx C'_D {{\bf
x}\over|{\bf x}|^D } \int d{\bf x}' \left[1- \exp\left(-\beta V({\bf
x}')\right)\right] \label{eqas}
\end{equation}
when the integral $F$ defined in equation (\ref{intf}) is finite.  For
$F$ to be finite we must have that $V(r)$ decays quicker than $1/r^D$
to ensure convergence as $r\to \infty$. In this case we see that
equation (\ref{eqas}) then implies that $g(r) \sim 1/r^{D-1}$ for
large $r$.

The boundary conditions on $g$ are that $g(r) \to 0$ as $r\to \infty$
(for $D\ge 2$) from the previous discussion and, if the potential $V$
is sufficiently well behaved, we expect that $g(0)$ is finite. In
greater than one dimension equation ({\ref{eqg}) is difficult to solve
explicitly. However for the case of soft sphere potentials of the form
$V_s(r) = \epsilon$ for $r <a$ and $V_s(r) = 0$ for $r > a$, one can
solve equation (\ref{eqg}) explicitly. In this case it is useful to
rewrite equation (\ref{eqg}) as
\begin{equation}
\fl {d\over dr}\left( \exp\left(-\beta V(r)\right) r^{D-1} {d\over dr}
g \right) + \exp\left(-\beta V(r)\right) r^{D-3} (1-D) g - r^{D-1}
{d\over dr}\exp\left(-\beta V(r))\right) = 0~~.
\end{equation}
This gives
\begin{equation}
{d\over dr}\left( r^{D-1} {d\over dr} g\right) + r^{D-3} (1-D) g = 0~~,
\label{eqgh}
\end{equation} 
for all $r\neq a$, supplemented by the conditions for matching the
discontinuity at $r=a$ which are
\begin{eqnarray}
& g(a^-) = g(a^+) \\ & a^{D-1}( {dg\over dr}(a^+) - \exp(-\beta
\epsilon){dg\over dr} (a^-)) - a^{D-1}(1-\exp(-\beta \epsilon)) = 0~~.
\end{eqnarray}
The general solution to equation (\ref{eqgh}) is
\begin{equation}
g(r) = {A\over r^{D-1}} + Br~~,
\end{equation}
and the boundary conditions at $r=0$ and as $r\to \infty$ along with
the continuity and jump condition across $r=a$ give
\begin{eqnarray}
g(r) &=& -{1-\exp(-\beta \epsilon)\over D-1 + \exp(-\beta \epsilon)}\
r\ \ \ \ \ \ \ \ \ \ \ \ ; r <a \nonumber \\ &=& -{1-\exp(-\beta
\epsilon)\over D-1 + \exp(-\beta \epsilon)} \ a \ ({a\over r})^{D-1} \
; r >a .
\end{eqnarray}
We thus obtain
\begin{eqnarray}
{\kappa_e^{(1)}\over \kappa} &=& 1 + {\rho S_D\over D} a^{D-1}
\left(1-\exp(-\beta \epsilon)\right) g(a) \nonumber \\ &=& 1- {\rho
S_D a^D\over D} {\left( 1 - \exp(-\beta \epsilon)\right)^2 \over (D-1
+ \exp(-\beta \epsilon))} ~~.\label{kappa1.1}
\end{eqnarray}
In the limit of hard spheres, that is to say as $\epsilon \to \infty$,
we obtain
\begin{equation}
{\kappa_e^{(1)}\over \kappa} = 1- {\rho S_D a^D\over D(D-1)}~~.
\end{equation}
This result is divergent at all $\rho$ for $D=1$ as it is clear that
in the presence of hard scatterers the tracer particle will be
confined between its nearest right and left neighbour for all time and
thus will not diffuse.
\section{Self-similar renormalisation group schemes}
The basic idea of self-similar renormalisation group schemes is to
resum the original perturbation expansions available to us in a
physically motivated way. Consider the general problem of diffusion in
a random potential where the potential is parameterised by $\Lambda$
such a way that when $\Lambda = 0$ the potential vanishes and the
resulting diffusion is free. For example we could use a Fourier based
parameterisation where we write
\begin{equation}
\phi_{\Lambda}({\bf x}) = \int_{|q|<\Lambda} {d{\bf q}\over
(2\pi)^{D}} {\tilde \phi}({\bf q}) \exp(i{\bf q} \cdot {\bf x})~~,
\end{equation}
Now we define a perturbation 
$\delta\phi_{\Lambda}({\bf x})$ to the potential so that
\begin{equation}
\phi_{\Lambda + \delta \Lambda} = \phi_{\Lambda} +
\delta\phi_{\Lambda}~~.
\end{equation} 
In the case of the Fourier parameterisation we have therefore
\begin{equation}
\delta\phi_{\Lambda}({\bf x}) = \int_{\Lambda <|q|\Lambda + \delta
\Lambda} {d{\bf q}\over (2\pi)^{D}} {\tilde \phi}({\bf q}) \exp(i{\bf
q} \cdot {\bf x})~~,
\end{equation}
which is the usual slicing technique in Fourier space.

One then integrates out the high momentum component
$\delta\phi_\Lambda$ perturbatively to find an effective theory where
the effects of $\delta\phi_\Lambda$ have been taken into account on
the running diffusion constant denoted by $\kappa(\Lambda)$ and the
coupling to the remaining gradient field denoted by
$\lambda(\Lambda)$. This is a self-similarity ansatz which means only
keeping interactions that were in the original problem and hence the
only parameters which change are $\kappa$ and $\lambda$ (as well as
the potential which just has the term $\delta\phi_\Lambda$ removed).
One therefore has a running diffusion constant $\kappa(\Lambda)$ and a
running coupling to the gradient field $\lambda(\Lambda)$. The
effective diffusion constant is then given by $\kappa_e =\kappa(0)$.
In addition it can be shown by general arguments \cite{der} that
\begin{equation}
{\kappa(\Lambda)\over \lambda(\Lambda)}={\kappa\over \lambda} = T,
\label{eqnfdt}
\end{equation}
that is to say that the Einstein relation, or fluctuation dissipation
relation, is satisfied by the renormalised theory at each step of the
renormalisation. This renormalisation is only valid for the
low-momentum component of the remaining drift and a possible
improvement to the calculation here would be to functionally
renormalise the remaining field $\phi_\Lambda$, which amounts to
introducing new interactions generated by the renormalisation
procedure (this approach has been applied to diffusion in an
incompressible quenched Gaussian velocity field \cite{dean2001}). If
$\nabla \phi = {\bf u}$ were a constant applied field then $\lambda$
would be the local conductivity coupling the particle to this external
field.

We remark that one may verify from the results on the weak disorder
perturbation theory in section 2 that the Fourier space based
renormalisation group method does not reproduce the exact results for
scatterering generated potentials known in one and two dimensions. The
Fourier based renormalisation group does however reproduce known exact
results in one and two dimensions for Gaussian potentials \cite{rg}.
However the Fourier space slicing of the potential is  not 
only the only decompostion possible. It was noted that in the 
case of diffusion in a Gaussian
field a so called t-slicing renormalisation group procedure
\cite{tslice} gave identical results to the Fourier space
renormalisation group. In the case of anisotropic potentials these two
renormalisation group schemes give different but numerically very
close results.  The basic idea behind the t-slicing approach is to
write
\begin{equation}
\phi_\Lambda = (\Lambda -\delta \Lambda)^{1\over 2}\phi' + {\delta
\Lambda}^{1\over 2} \phi''~~,
\end{equation}
where $\phi'$ and $\phi''$ are two independent Gaussian fields with
the same correlation function as the original field $\phi$. In the
above prescription $\Lambda$ is integrated down from one to zero, by
perturbatively calculating, and averaging over, the effect of term
$\phi''$.  The corresponding flow equations for $\kappa(\Lambda)$ and
$\lambda(\Lambda)$ are then integrated. The physical argument for the
Fourier based renormalisation group scheme is that at length scales
$1/ \Lambda$ the effective diffusion constant and effective
`conductivity' $\lambda$ is determined by fluctuations in the field on
length scales shorter than this. In the case of the scatterers one can
apply a self similar renormalisation group approach based on thinning
out the density and which has the advantage of agreeing with known
exact results. Concretely, one assume that the effect of a density
$\rho$ of scatterers is to generate an effective diffusion process
described by an effective diffusion constant $\kappa(\rho)$ and
coupling $\lambda(\rho)$ on length scales bigger that the
inter-particle separation. We now throw in more scatterers of the same
type of small density $\delta \rho$. As $\delta\rho$ is small we may
use the results of the previous section along with the self-similarity
ansatz here to write
\begin{equation}
{\delta\kappa(\rho)\over \kappa(\rho)} = {\delta\rho S_D\over D}
\int_0^\infty r^{D-1} dr\ \left[{\partial\over \partial r}
\exp\left(-\beta V(r) \right)\right] g(r)~~,
\end{equation}
which integrates to give
\begin{equation}
\kappa(\rho)= \kappa\exp\left({\rho S_D\over D} \int_0^\infty r^{D-1}
dr\ \left[{\partial\over \partial r} \exp\left(-\beta V(r)
\right)\right] g(r)\right)~~. \label{rgrho}
\end{equation}
We see that the effect of the renormalisation group procedure is to
exponentiate the small $\rho$ expansion equation (\ref{smallrho}).
Immediately we see from its multiplicative form that the result
equation (\ref{smallrho}) agrees with the known results in one and two
dimensions referred to previously. Note it is important that the
scatterers are uniformly and independently distributed as the added
scatterers at each step have no information about those already
integrated out. Let us also note that on adding an additional $\delta
\rho$ of scatterers, the characteristic length between these
scatterers is $l^* \sim 1/{\delta \rho}^{1\over D}$ which is strictly
infinite as $\delta \rho \to 0$, hence a diffusing particle will, on
average, have acquired the effective diffusion constant $\kappa(\rho)$
and conductivity $\lambda(\rho)$ before encountering a newly added
scatterer.  For a system of soft spheres we have in the RG approach
\begin{equation}
{\kappa_e\over \kappa} = \exp\left[- {\rho S_D a^D\over D} {\left( 1 -
\exp(-\beta \epsilon)\right)^2 \over (D-1 + \exp(-\beta \epsilon))}
\right]~~.
\label{kssrg}
\end{equation}
Despite its agreement with known exact results in one and two
dimensions the expression equation (\ref{rgrho}) will not be exact in
general.  Indeed, if one considers a system of hard spheres one
expects that there is a percolation transition at a density $\rho_c$
above which the diffusion constant vanishes, the formula equation
(\ref{kssrg}) is clearly insensitive to any such transition; we see
that in the hard sphere limit $\epsilon\to \infty$ that $\kappa_e =
\exp\left(-\rho S_D a^D/(D (D-1)) \right)$. It is inevitable that this
approximate result will be inaccurate at high densities.  In the
following section we shall compare the results obtained by the various
techniques used here with numerical simulations.

A related problem to the one studied here is the problem of
calculating the effective diffusivity of a medium with locally varying
diffusivity. The diffusion equation for this system is
\begin{equation}
{\partial p \over \partial t} = \nabla \cdot \left(\kappa({\bf
x})\nabla p\right)~~.
\label{eqdiff}
\end{equation}
The effective diffusion constant in this model also corresponds to the
effective dielectric permeability of a medium with local permeablility
$\kappa(\bf x)$ and thus we shall denote it by $\kappa_e^{(p)}$. In
the case where
\begin{equation}
\kappa({\bf x} ) = \kappa \exp(-\beta \phi({\bf x}))~~,
\end{equation}
one can relate the diffusion constant for a particle diffusing in the
potential $\phi$ at inverse temperature $\beta$ with the bare
diffusion constant $\kappa$, to that of the effective diffusion
constant of the process defined by equation (\ref{eqdiff}) by the
remarkably simple formula
\begin{equation}
\kappa_e^{(p)} = {\kappa_e {\overline \kappa}\over \kappa}~~.
\label{kappar}
\end{equation}
We denote by ${\overline\kappa}$ the spatial average of the local
diffusivity or permeability
\begin{equation}
{\overline\kappa} = {1\over {\cal V}}\int_V d{\bf x}\ \kappa({\bf x})~~.
\label{kappal}
\end{equation} 
This result was shown in \cite{der} via an algebraic approach but it
has a simple probabilistic derivation which we include in the
Appendix.  In the case considered here, of uniform scatterers, we have
\begin{equation}
{\overline\kappa} = \kappa \exp\left[\rho\int d{\bf x} ( \exp(-\beta
V) -1)\right].
\label{kappav}
\end{equation}
The study of the effective dielectric constants of a mixture of
dielectric spheres suspended in an otherwise uniform medium dates back
to Maxwell \cite{max}. If the spheres have dielectric constant
$\kappa'$ and the background medium has dielectric constant $\kappa$,
then the local dielectric constant may be written as
\begin{equation}
\kappa({\bf x}) = \kappa \exp\left(-\phi({\bf x})\right)~~,
\end{equation}
where
\begin{equation}
\phi = \sum_i V({\bf x} -{\bf x}_i)~~,
\end{equation}
with $V(r)$ given by the soft sphere potential: $V(r)=
\ln(\kappa/\kappa')$ for $r < a$ and $V(r)= 0$ for $r >a$, with $a$
the radius of the spheres.  This formulation clearly works when the
positions of the spheres are chosen to be non-overlapping. However, to
order $\rho$ the spheres do not overlap and thus from the calculations
developed here, using equation (\ref{kappa1.1}) with $\epsilon =
\ln(\kappa/\kappa')$, equation (\ref{kappar}) and equation
(\ref{kappav}) to order $\rho$, we find
\begin{eqnarray}
\fl {\kappa_e^{(p)} \over \kappa} &\approx& (1 - {\rho S_D a^D\over
D}{\left(1 - {\kappa'\over \kappa}\right)^2 \over \left(D-1 +
{\kappa'\over \kappa}\right)})\times\left(1 + {\rho S_D a^D\over
D}({\kappa'\over \kappa}-1)\right) \nonumber \\ &\approx& 1 + {c D
(\alpha -1)\over (D-1) + \alpha}~~,
\end{eqnarray}
where $c= \rho S_D a^D/D$ is the volume fraction of the spheres and
$\alpha = \kappa'/\kappa$. In three dimensions this result gives
\begin{equation}
{\kappa_e^{(p)} \over \kappa} \approx 1 + {3c (\alpha -1)\over 2 +
\alpha}~~,
\end{equation}
which agrees with Maxwell's result, which he gives as \cite{max}
\begin{equation}
{\kappa_e^{(p)} \over \kappa} = {\alpha + 2 + 2c(\alpha-1)\over \alpha
+ 2 - c(\alpha-1)}~~,
\label{maxe}
\end{equation}
to first order in $c$.  Note that the Maxwell result is only accurate
to order $c$ and equation (\ref{maxe}) is a physically motivated
resummation of the order $c$ result, it is sometimes referred to in
the literature as the Clausius-Mosotti equation.  The correction at
order $c^2$ was calculated many years later by Jeffrey
\cite{jeff}. Despite the fact that the calculation carried out in this
paper is for spheres which can overlap, it is interesting to compare
the full renormalisation group result for overlapping spheres with
results obtained for the effective dielectric constant for a system of
non-overlapping spheres via numerical calculation. The renormalisation
group calculation for overlapping spheres gives
\begin{equation}
\kappa_e^{(p)} = \kappa\exp\left(3c \ {\alpha -1\over \alpha + 2}
\right)~~.
\label{srg}
\end{equation}

Brady and Bonnecaze \cite{bobr} numerically computed the effective
dielectric constant for non-overlapping dielectric spheres in a
background medium. In Tables (1-3) we compare their results with
equation (\ref{srg}), the Maxwell formula equation (\ref{maxe}) and
Jeffrey's result which is correct to $O(c^2)$.  For $\alpha = \infty$
and $\alpha = 10$, the RG result is surprisingly close to the
numerically obtained values and performs better than the Maxwell
formula and the $O(c^2)$ result. For $\alpha=0.01$ it performs less
well than the two other results.

\begin{table}
\caption{\label{alinf} Calculations of $\kappa_e^{(p)}/\kappa$ for
$\alpha = \infty$ }
\begin{tabular}{@{}lllll}
\br c& numerical & $O(c^2)$ result & Maxwell & RG\\ \mr 0.00 & 1.000 &
1.000 & 1.000 & 1.000 \\ 0.10 & 1.352 $\pm$ 0.030& 1.345 & 1.333 &
1.349 \\ 0.20 & 1.821 $\pm$ 0.081& 1.780 & 1.750 & 1.822 \\ 0.30 &
2.529 $\pm$ 0.168& 2.306 & 2.236 & 2.459 \\ 0.40 & 3.590 $\pm$ 0.230&
2.923 & 3.000 & 3.320 \\ 0.50 & 4.967 $\pm$ 0.339& 3.628 & 4.000 &
4.482 \\ \br
\end{tabular}
\end{table}

The effective potential for the scatterers in the potential diffusion
problem was $\epsilon = -\ln(\alpha)$. Hence if $\alpha > 1$ then the
potential $\epsilon$ is negative and thus attractive to a particle
diffusing in that potential. The diffusion of the particle is clearly
slowed down by trapping inside the scatterers, and it appears that the
RG is picking up this dominant effect, correlations between scatterers
give a small effect.
\begin{table}
\caption{\label{alinfb} Calculations of $\kappa_e^{(p)}/\kappa$ for
$\alpha = 10$ }
\begin{tabular}{@{}lllll}
\br c& numerical & $O(c^2)$ result & Maxwell & RG\\ \mr 0.00 & 1.000 &
1.000 & 1.000 & 1.000 \\ 0.10 & 1.247$\pm$ 0.011 & 1.240 & 1.243 &
1.252 \\ 0.20 & 1.545$\pm$ 0.028 & 1.511 & 1.529 & 1.568 \\ 0.30 &
1.944$\pm$ 0.046 & 1.812 & 1.871 & 1.964 \\ 0.40 & 2.443$\pm$ 0.046 &
2.144 & 2.286 & 2.460 \\ 0.50 & 3.080$\pm$ 0.053 & 2.507 & 2.800 &
3.080 \\ \br
\end{tabular}
\end{table}

If we consider the case where $\alpha < 1$, then the potential
$\epsilon$ is positive and repels the tracer. The slowing down of the
diffusion is now mainly induced by another, much more subtle, effect:
a memory effect which is much more sensitive to the correlations
between the scatterers.
\begin{table}
\caption{\label{alinfc} Calculations of $\kappa_e^{(p)}/\kappa$ for
$\alpha = 0.01$ }
\begin{tabular}{@{}lllll}
\br c& numerical & $O(c^2)$ result & Maxwell & RG\\ \mr 0.00 & 1.000 &
1.000 & 1.000 & 1.000 \\ 0.10 & 0.859$\pm$ 0.004 & 0.858 & 0.859 &
0.862 \\ 0.20 & 0.729$\pm$ 0.005 & 0.727 & 0.731 & 0.744 \\ 0.30 &
0.600$\pm$ 0.007 & 0.608 & 0.614 & 0.641 \\ 0.40 & 0.504$\pm$ 0.010 &
0.501 & 0.506 & 0.554 \\ 0.50 & 0.406$\pm$ 0.012 & 0.455 & 0.407 &
0.478 \\ \br
\end{tabular}
\end{table}

\section{Simulation Results}
The values of the effective diffusion constant as predicted by
Equation (\ref{kappa1.1}) and (\ref{kssrg}) have been compared to the
effective diffusion constant measured in Monte Carlo simulations of
two dimensional systems of soft spheres. The system is a square of
linear size $L=20$, with the spheres of radius, as defined by Equation
(\ref{kappa1.1}), $a=1$ uniformly distributed in the area. As usual
periodic boundary conditions are imposed. The simulations consisted of
$2000$ independent runs of a single tracer particle during a total
time $t_{max}=10^5$, with an elementary time increment $\delta
t=0.0025$.  Each particle was evolved in an independent realisation of
the disorder.  The value of $\kappa_e$ was evaluated by fitting the
ensemble average of ${\bf X}_t^2/t$ to $4\kappa_e + A\ln(t)/t$ at late
times (see Equation (\ref{eqcorrs})). The errors were then estimated
from the typical late time fluctuation of the ensemble averaged curve
about the fitting curve.  In Figure (\ref{fig:-1}) $\kappa_e$ is
plotted as a function of $T$, for $\rho=0.35$ and $\epsilon=-1$, {\em
i.e.} a case where the scatterers attract, and tend to trap, the
tracer. The agreement with the RG calculation is extremely good, the
RG calculation correctly predicts the behaviour of $\kappa_e$ across a
range where it varies almost by a factor of $5$. In this case the
accuracy of the RG is comparable to that when it is applied to the
problem of diffusion in a Gaussian random potential in three
dimensions \cite{rg}.

In Figure (\ref{fig:1}) $\kappa_e$ is plotted as a function of $T$,
for $\rho=0.55$ and $\epsilon=1$, thus the scatterers repel the
tracer. In this case the RG works notably less well and the numerical
curves appear to be bounded above by the RG and bounded below by the
order $\rho$ result.

Finally in Figure (\ref{fig:infty}) $\kappa_e$ is plotted as a
function of $\rho$ for $\epsilon=\infty$, the case of hard sphere
scatterers. Here the perturbative order $\rho$ result performs
extremely well and is superior to the RG result. This may be expected
as the ratio $\lambda_0/\kappa_0$ in this case is effectively infinite
and the renormalisation in the zero temperature case may not make
sense.

The comparisons here are in accordance with those seen in the last
section, the RG seems to perform well when the slowing down of the
diffusion is due to trapping centres. It performs significantly less
well in the case of repulsive scatterers where the slowing down of the
diffusion is due to more subtle correlation effects and presumably
needs a higher order treatment of terms in the expansion in $\rho$.

\begin{figure}
\begin{center}
\epsfxsize=0.7\hsize\epsfbox{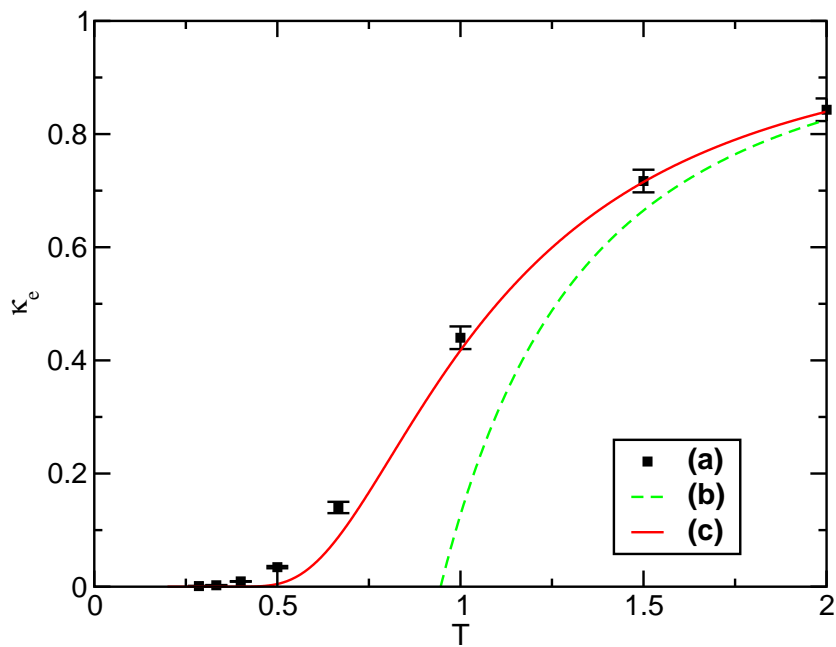}
\end{center}
\caption{\label{fig:-1} Effective diffusion constant as a function of
the temperature, for $\epsilon=-1$ and $\rho=0.35$: from numerical
simulations (a), low density approximation (b) and density RG (c).}
\end{figure}

\begin{figure}
\begin{center}
\epsfxsize=0.7\hsize\epsfbox{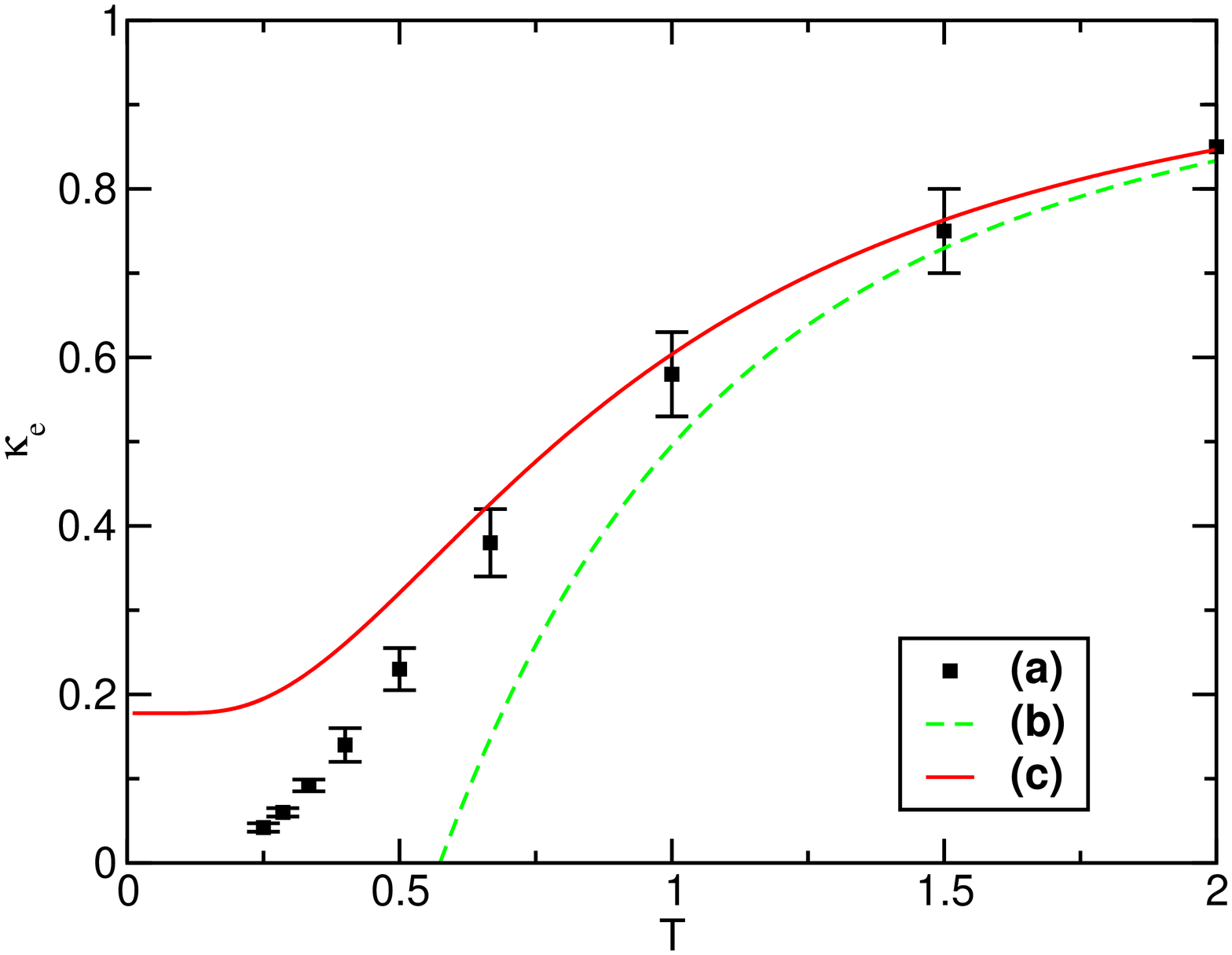}
\end{center}
\caption{\label{fig:1} Effective diffusion constant as a function of
the temperature, for $\epsilon=1$ and $\rho=0.55$: from numerical
simulations (a), low density approximation (b) and density RG (c). }
\end{figure}
\begin{figure}

\begin{center}
\epsfxsize=0.7\hsize\epsfbox{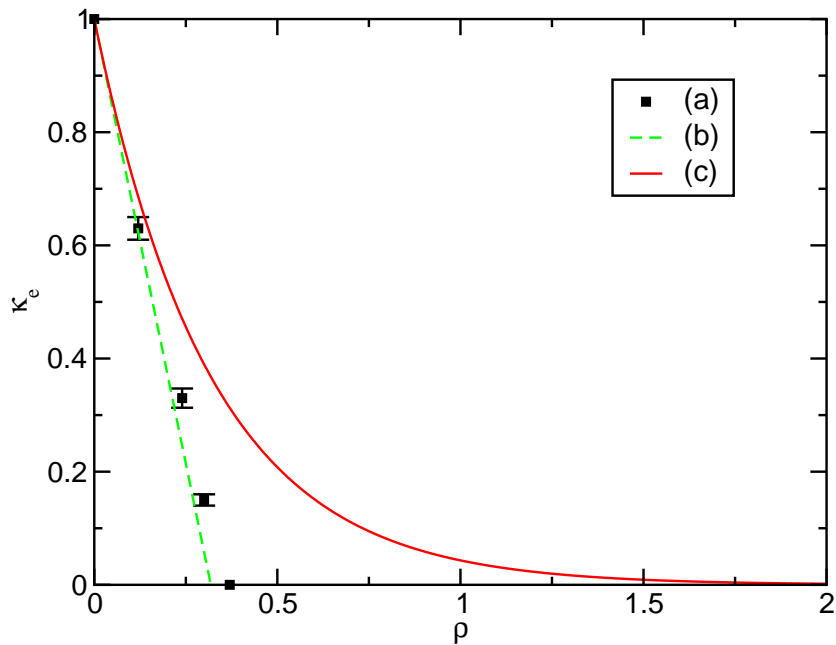}
\end{center}
\caption{\label{fig:infty} Effective diffusion constant as a function
of the density for $\epsilon=\infty$ (overlapping hard spheres): from
numerical simulations (a), low density approximation (b) and density
RG (c).}
\end{figure}

\section{Conclusion}
In this paper we have considered the problem of a Langevin process
diffusing in a quenched random potential. The random potential is
generated by uniformly and independently distributed scatterers. We
developed a weak coupling or high temperature perturbation theory
which can be used to determine the validity of the Gaussian
approximation to the random field distribution. As expected from the
central limit theorem, one of the conditions needed to validate the
Gaussian approximation is that there is a high density of scatterers
so that the potential at each point is the sum of a large number
random variables. However this condition is only sufficient at high
temperatures. Physically this can be understood as follows: even
though, due to the central limit theorem, at a randomly chosen point
the potential will be Gaussian, at low temperatures the diffusion will
be dominantly near the minima of the potential which by definition
belong to the tails of the distribution where the central limit
theorem does not apply as they are at the extreme points of the
distribution.  We have also seen that the self-similar RG resummation
of the weak disorder expansion does not reproduce known exact
results. In addition the weak disorder expansion is destined to fail
for singular potentials, where the potential contains a divergence. We
thus formulated the self-similar RG resummation in terms of a
decimation or thinning procedure in terms of the particle
density. Direct and indirect comparison with numerical results for
hard and soft sphere potentials show that the density based RG
resummation gives reasonable predictions for the bulk diffusivity and
agrees with known exact results. It appears to work best in cases
where the scatterers have a potential which is attractive to the
tracer and thus act as trapping centres. In the case where the
scatterers are not trapping but are repelling, diffusion is again
slowed down but by a more subtle correlation mechanism, this effect is
less well described by the RG result.
    
\section*{Acknowledgements}

A. L. acknowledges support from the grant HPRN-CT-2002-00319 of the
European STIPCO network.  \appendix

\section{Expression for $\kappa_e$}
Here we derive the representation equation (\ref{krep}) for the
effective diffusion constant. The basic idea of the proof in much more
mathematical terms may be found in the book of Spohn \cite{spohn}.
Consider the process in equilibrium in a large volume ${\cal V}$ at the time
$t=0$. Let the initial position be ${\bf X}_0$ distributed according
to the equilibrium measure $p_{eq}({\bf x}_0) = \exp(-\beta \phi({\bf
x}_0))/Z$ where $Z$ is the one particle partition function normalising
the distribution. Integrating the stochastic differential equation
(\ref{eqlan}) between $0$ and $t$ we obtain
\begin{equation}
{\bf X}_t -{\bf X}_0 = \sqrt{\kappa} {\bf B}_t - \lambda \int_0^t ds\
\nabla \phi({\bf X}_s)~~ ,
\end{equation}
where ${\bf B}_t$ is a standard $D$ dimensional Brownian motion with
$\langle {\bf B}^2_t\rangle =2 D t$.  Squaring the above and taking
its average yields yields
\begin{equation}
\fl 2D \kappa t = \langle ( {\bf X}_t -{\bf X}_0 )^2 \rangle +
2\lambda \langle ( {\bf X}_t -{\bf X}_0 )\cdot \int_0^t ds\ \nabla
\phi({\bf X}_s) \rangle + 2\lambda^2\langle \int_0^t ds \int_0^sds'\
\nabla \phi({\bf X}_s) \cdot \nabla \phi({\bf X}_{s'})~~.  \rangle
\end{equation}
Defining a time dependent diffusion constant of $\kappa^*(t) =\langle(
{\bf X}_t -{\bf X}_0 )^2\rangle/2Dt$, so that $\lim_{t\to \infty}
\kappa^*(t) = \kappa_e$ this yields
\begin{equation}
\fl \kappa^*(t) = \kappa - {\lambda\over D t} \langle ( {\bf X}_t
-{\bf X}_0 )\cdot \int_0^t ds\ \nabla \phi({\bf X}_s) \rangle -
{\lambda^2\over Dt}\langle \int_0^t ds \int_0^s ds'\ \nabla \phi({\bf
  X}_s) \cdot \nabla \phi({\bf X}_{s'}) \rangle~~. \label{eqa1}
\end{equation}
Using the property of detailed balance, which ensures time translation
invariance at equilibrium, we have for any two functions $A$ and $B$
that
\begin{equation}
\langle A({\bf X}_t) B({\bf X}_s)\rangle = \langle A({\bf X}_{t-s})
B({\bf X}_0)\rangle~~,
\end{equation}
and the symmetry relation
\begin{equation}
\langle A({\bf X}_t) B({\bf X}_s)\rangle = \langle B({\bf X}_t) A({\bf
X}_s) \rangle ~~.
\end{equation}
Using this it is easy to verify that the second term on the right hand
side of equation (\ref{eqa1}) is zero.  The third term may be
simplified using
\begin{eqnarray}
  I(t) &=& \langle \int_0^t ds \int_0^s ds'\ \nabla \phi({\bf X}_s)
\cdot \nabla \phi({\bf X}_{s'})\rangle \nonumber \\ &=&
\langle\int_0^t ds \int_0^s du\ \nabla \phi({\bf X}_u) \cdot \nabla
\phi({\bf X}_0)\rangle \nonumber \\ &=& \int d{\bf x} d{\bf
x}_0\int_0^t ds \int_0^s du\ \nabla \phi({\bf x}) p({\bf x},{\bf
x}_0,u)\cdot \nabla \phi({\bf x}_0) p_{eq}({\bf x}_0)~~,
\end{eqnarray}
where $p({\bf x},{\bf x}_0,u)$ is the transition density to go from
${\bf x}_0$, to ${\bf x}$ in time $u$ and we have used the fact that
the density for the starting point ${\bf x}_0$ is the equilibrium
density.  We now use the fact that formally
\begin{equation}
p({\bf x},{\bf x}_0,u) = \exp(-u H({\bf x},{\bf x}_0))~~,
\end{equation}
to obtain
\begin{eqnarray}
I(t) &=& \int d{\bf x} d{\bf x}_0 \int_0^t ds \nabla \phi({\bf x})
H^{-1} (1- \exp(-s H))\cdot \nabla \phi({\bf x}_0) p_{eq}({\bf x}_0)
\nonumber \\ &=& t \int d{\bf x} d{\bf x}_0 \nabla \phi({\bf x})
H^{-1}\cdot \nabla \phi({\bf x}_0) p_{eq}({\bf x}_0) \nonumber \\ &-&
\int d{\bf x} d{\bf x}_0 \nabla \phi({\bf x}) H^{-2}(1-\exp(-t
H))\cdot \nabla \phi({\bf x}_0) p_{eq}({\bf x}_0)~~,
\label{timec}
\end{eqnarray}
where in the above the notation of multiplication by $H^{-1}$
corresponds to operator composition. Using the definition of $G'$ and
putting this altogether we find, keeping the dominant terms in the
large $t$ expansion, that
\begin{equation}
\kappa_e = \kappa - {\lambda^2 \over D} \int d{\bf x} d{\bf x}_0
\nabla \phi({\bf x}) \cdot G({\bf x}, {\bf x}_0) \nabla \phi({\bf
x}_0) p_{eq}({\bf x}_0)~~,
\end{equation}
which is the desired result. Note that all the above derivation
depends on the fact that $H$ is a positive definite operator. In a
finite volume this is clearly not the case as there is an equilibrium
distribution.  The result should be understood for long times but such
that the system has not encountered the boundary of ${\cal V}$. If in the
above the limit $t\to \infty$ is taken before $|{\cal V}| \to \infty$ one
finds that $\kappa_e = 0$ which is to be expected for a finite
system. A more detailed discussion of this subtlety can be found in
\cite{der}.

Another interesting consequence of equation (\ref{timec}) is that
using the zeroth order (or free diffusion) result for $H^{-2}$ and
$\exp(-t H)$ in the finite time correction in equation (\ref{timec}),
we obtain the expected finite time correction \cite{spohn} to the long
time diffusive behaviour, {\em i.e.}
\begin{eqnarray}
\langle {\bf X}^2_t\rangle &=& 2\kappa_e t + O(\sqrt{t})\ \ \ \ ;D=1
\\ &=& 4\kappa_e t + O(\ln(t)) \ \ ;D=2 \\ &=& 2D\kappa_e t + O(1) \ \
\ \ ; D\geq 3~~.
\label{eqcorrs}
\end{eqnarray}

\section{Relationship between $\kappa_e$ and $\kappa_e^{(p)}$}

The diffusion equation equation (\ref{eqdiff}) corresponds to a
process which in the Ito prescription of the stochastic calculus obeys
\begin{equation}
d{\bf X}=\sqrt{\kappa\exp(-\beta \phi({\bf X}))} \eta(t) dt -
\kappa\beta\exp(-\beta \phi({\bf X})) \nabla\phi({\bf X}) dt~~
\label{sdk}
\end{equation}
where $\kappa({\bf x}) = \kappa\exp(-\beta \phi({\bf x}))$.  We bear in
mind that the equilibrium distribution of this process can readily be
seen from equation (\ref{eqdiff}) to be flat, {\em i.e} $p_{eq}({\bf
x}) = 1/{\cal V}$. Now define a random time variable
\begin{equation}
\tau = \int_0^t ds\ \exp(-\beta\phi({\bf X}_s))
\end{equation}
which is clearly a monotonic proper, though random, change of time
scales.  Clearly we may write equation (\ref{sdk}) as
\begin{equation}
d{\bf X}=\sqrt{\kappa\exp(\beta \phi({\bf X}))} \eta(t) d\tau -
\lambda \nabla\phi({\bf X}) d\tau~~ ,\label{sdk2}
\end{equation}
and we have used $\kappa \beta = \lambda$.  Formally we may write
\begin{equation}
\eta(t) = \sigma/\sqrt{dt}~~,
\end{equation}
where $\sigma$ is a normal random variable of zero mean and variance
2.  We can therefore write equation (\ref{sdk2}) entirely in terms of
the time variable $\tau$
\begin{equation}
d{\bf X}=\sqrt{\kappa} \eta(\tau) d\tau - \lambda \nabla\phi({\bf X})
d\tau~~.
\end{equation}
This equation is just that for a particle diffusing in the potential
$\phi$ at inverse temperature $\beta$. Thus by definition
\begin{equation}
\langle {\bf X}^2_\tau \rangle = 2 \kappa_e D \tau
\end{equation}
but also
\begin{equation}
\langle {\bf X}^2_t \rangle = 2 \kappa_e^{(p)} D t~~.
\end{equation}
Comparing the two expressions which must be the same one finds
\begin{eqnarray}
\kappa_e^{(p)} &=& \kappa_e {\tau\over t} \nonumber \\ &=& \kappa_e
                {1\over t} \int_0^t ds \ \exp(-\beta\phi({\bf X}_s))~~.
\end{eqnarray}
We now use the fact that ${\bf X}$ has a flat equilibrium distribution
to give that (for large times)
\begin{equation}
 {1\over t} \int_0^t ds \ \exp(-\beta\phi({\bf X}_s)) \to {1\over {\cal V}}
\int_{\cal V} d{\bf x} \exp(-\beta\phi({\bf x}))~~,
\end{equation}
which thus leads to the desired result equation (\ref{kappar}).

\end{document}